\documentclass{article}

\PassOptionsToPackage{numbers, compress}{natbib}

\usepackage[preprint]{neurips_2019_ml4ps}

\usepackage[utf8]{inputenc}   
\usepackage[T1]{fontenc}      
\usepackage{hyperref}         
\usepackage{url}              
\usepackage{booktabs}         
\usepackage{amsmath}
\usepackage{amsfonts}         
\usepackage{nicefrac}         
\usepackage{microtype}        
\usepackage[pdftex]{graphicx} %
\usepackage{array}
\usepackage{multirow}

\usepackage{feynmp-auto}
\usepackage{subcaption}
\usepackage{siunitx}

\usepackage{bm}

\newcommand{\specialcellbold}[2][c]{%
  \bfseries
  \begin{tabular}[#1]{@{}l@{}}#2\end{tabular}%
}

\newcommand{\bz}{\boldsymbol{z}}
\newcommand{\bx}{\boldsymbol{x}}

\title{Variational Autoencoders for Generative Modelling of Water Cherenkov Detectors}

\author{
  Abhishek Abhishek \\
  University of Manitoba/TRIUMF \\
  \texttt{abhishek@myumanitoba.ca} \\
\And
  Wojciech Fedorko \\
  TRIUMF \\
  \texttt{wfedorko@triumf.ca} \\
\And
  Patrick de Perio \\
  TRIUMF \\
  \texttt{pdeperio@triumf.ca} \\
\And
  Nicholas Prouse \\
  TRIUMF \\
  \texttt{nprouse@triumf.ca} \\
\And
  Julian Z. Ding \\
  University of British Columbia/TRIUMF \\
  \texttt{julianzding@gmail.com}
}

\begin{document}

\maketitle

\begin{abstract}
  Matter-antimatter asymmetry is one of the major unsolved problems in physics that can be probed through precision measurements of charge-parity symmetry violation at current and next-generation neutrino oscillation experiments. In this work, we demonstrate the capability of variational autoencoders and normalizing flows to approximate the generative distribution of simulated data for water Cherenkov detectors commonly used in these experiments. We study the performance of these methods and their applicability for semi-supervised learning and synthetic data generation.

\end{abstract}

\section{Introduction}
	We currently cannot explain the observed matter-antimatter asymmetry in the universe. Neutrino oscillations~\citep{SNOevidence,fukuda1998evidence} may exhibit significant charge-parity symmetry violation (CPV)~\citep{Sakharov_1991}, which may lead to the answer~\citep{FUKUGITA198645leptogenesis}. These can be measured in experiments such as T2K~\citep{t2k}, which produce a muon neutrino ($\nu_\mu$) or antineutrino beam directed towards a far detector where the oscillation signal is electron neutrinos ($\nu_e$) and antineutrinos, respectively.

The planned next generation Hyper-Kamiokande (Hyper-K)~\citep{abe2018hyper} far detector and NuPRISM~\citep{bhadra2014letter} intermediate water Cherenkov detector (IWCD) are multi-kilotonne water tanks surrounded by $\mathcal{O}(10k)$ 20'' photomultiplier tubes (PMTs) or multi-PMT (mPMT) modules, consisting of 3'' PMT arrays. They detect single photons of Cherenkov light from muons ($\mu^-$) and electrons ($e^-$) produced in $\nu_\mu$ and $\nu_e$ interactions with water, respectively. These are easily classified by traditional likelihood ratio-based methods~\cite[e.g.][]{fitqun} due to the electromagnetic shower induced by the much lighter electron, causing a fuzzier Cherenkov light ring projected onto the PMTs. These methods, however, have limited discriminative power between an $e^-$ and a high energy photon ($\gamma$) produced by a $\nu_\mu$ without the corresponding $\mu^-$. Such a photon can produce an $e^-$ and $e^+$ pair, which then also shower resulting in a very similar ring as an $e^-$ event. Such misclassified events constitute a significant and poorly understood background to the $\nu_e$ signal~\citep{T2K:NCgamma, T2KLatestShort, T2KLatestLong}, with no existing experimental constraint.

As in existing WC detectors such as Super-Kamiokande (Super-K)~\citep{fukuda2003super},
 1) accurate first-principles modelling of detector effects, such as varying water conditions or PMT responses, remains a significant challenge. Furthermore, 2) due to the improved granularity of IWCD and potentially the Hyper-K detector, it may be possible to detect the slight difference in Cherenkov light emission between an $e^-$ from a $\nu_e$ and $e^-/e^+$ pair from a $\gamma$. Thus, the implementation and development of modern machine learning methods can 1) enable training directly on data to mitigate experimental uncertainties, and 2) maximize the rejection of $\gamma$ events to limit exposure to unconstrained theoretical uncertainties. Both developments are expected to enhance the sensitivity of Hyper-K to CPV and other phenomena such as proton decay~\citep{protondecay} and multi-messenger astronomical events~\citep[e.g.][]{SN1987a,SNEWS,IceCube}.

In this work we study the capability of variational autoencoders (VAEs) and their extensions: normalizing flows (NF) to learn the data generating distribution and benchmark the performance of VAEs in semi-supervised training on simulated IWCD datasets. In addition, we study the capability of VAEs to generate synthetic samples.

\section{Methodology}
	\subsection{Simulation and Data Preprocessing}

    Water Cherenkov Simulation (WCSim)\footnote{https://github.com/WCSim/WCSim} is a GEANT4~\citep{agostinelli2003geant4} and ROOT \citep{brun1997root} based Monte Carlo (MC) software package. It was used to generate 3 million events each of $e^{-}$, $\mu^{-}$, and $\gamma$ particles in the IWCD. For this initial study, the $e^{-}$, $\mu^{-}$ initial positions and $\gamma$ pair production positions were fixed at the center of the detector. In order to use convolutional neural networks (CNNs) as our primary NN architecture, the top and bottom of the detector cylinder were ignored and the particle directions were constrained to be perpendicular to the detector wall. The azimuthal angles and the energies of the particles were uniformly varied between $0$ to $2\pi$ and $0$ to $1000$\ MeV. The detector wall is instrumented with $16 \times 40$ mPMT ``pixels'' and the resulting data is structured as an image with each pixel containing $19$ channels corresponding to the light intensity measured by each $3"$ PMT.

\subsection{Semi-supervised and Unsupervised Deep Generative Models}

    \textbf{Variational Autoencoders (VAEs)}~\cite{kingma2014stochastic} are a powerful class of latent generative models that maximize an evidence lower bound (ELBO) to the intractable log-likelihood of the data, log $p_{\theta}(\bx)$:
    
    \begin{equation}
      \label{eq:VAE}
      \log{p_{\theta}(\bx)} \geq E_{\bz \sim q_{\phi}(\bz|\bx)}[\log{p_{\theta}(\bx|\bz)}] - D_{KL}[q_{\phi}(\bz|\bx)||p(\bz)] = -\mathcal{J}
    \end{equation}
    
    where $D_{KL}$ is the Kullback-Leibler divergence, $p(\bz)$ is the prior distribution over the latent variables $\bz$,  $q_{\phi}(\bz|\bx)$ is the variational posterior distribution, and  $p_{\theta}(\bx|\bz)$ is the conditional generative distribution. Generally, $p_{\theta}(\bx|\bz)$ and $p(\bz)$ are defined as factorized Gaussian distributions $\mathcal{N}(\bz|\boldsymbol{\mu}_{\phi}(\bx),diag(\boldsymbol{\sigma}^{2}_{\phi}(\bx)))$ and $\mathcal{N}(0,I)$, respectively. $q_{\phi}(\bz|\bx)$ and $p_{\theta}(\bx|\bz)$ are parameterized using neural networks with parameters $\theta$ and $\phi$, learned through minimization of $\mathcal{J}$ by gradient descent methods.The expectation over the conditional generative distribution $E_{\bz \sim q_{\phi}(\bz|\bx)}[\log{p_{\theta}(\bx|\bz)}]$ corresponds to the fidelity of the reconstruction. Since the input variables $\bx$ are continuous and considered Gaussian distributed, this term is replaced by negative Mean Squared Error (MSE) loss.

    \textbf{Normalizing Flows (NFs)}~ \cite{rezende2015variational} were developed to address a key limitation of VAEs where, even in the asymptotic regime, one is unable to recover the true posterior distribution $p_{\theta}(\bz|\bx)$ due to the simple form of $q_{\phi}(\bz|\bx)$. In normalizing flows, a random variable $z_{0}$ with an initial probability density $q_{0}(z_{0})$ is transformed into another random variable $z_{K}$ with a probability density $q_{K}(\bz_{K})$ through a sequence of $K$ smooth, invertible mappings, $f$:
    \begin{equation}
        \label{eq:Flow}
        \bz_{K} = f_{K} \circ f_{K-1} \circ ... \circ f_{2} \circ f_{1}(\bz_{0}).
    \end{equation}
    
    In planar normalizing flows, these mappings have the form:
    \begin{equation}
        \label{eq:PNF}
        f_k(\bz) = \bz_{k-1} + \boldsymbol{u}_{k}h(\boldsymbol{w}_{k}^{T}\bz_{k-1} + \boldsymbol{b}_k),  \hspace{1mm} k \in [1,K],
    \end{equation}
     where $\boldsymbol{u}$, $\boldsymbol{w}$, and $\boldsymbol{b}$  are the free parameters of the flow and $h$ is a smooth element-wise non-linearity.

\subsection{Model architecture}

    The architecture of a VAE can be conceptually divided into an encoder, a ``bottleneck'', and a decoder. We employed a simple LeNet \cite{lecun1998gradient} inspired CNN as the encoder and a symmetric architecture comprising of transposed convolution layers as the decoder. The encoder comprised of four $3 \times 3$ and two $4 \times 4$ convolution layers with the number of channels per pixel successively increasing from $19$ to $128$. ReLU nonlinearities were used after each layer. Finally, $4$ fully connected layers were used to successively compress the feature vectors to the size of the latent vectors. The simple choice of the CNN architecture allowed for focus on the development of the generative models rather than the NN architectures.

\section{Empirical Evaluations and Results}
	The 9 million event dataset is split into 80\%-10\%-10\% for train, validation, and test subsets, respectively. During the training process, the model with the lowest loss on a validation mini-batch  is retained for further evaluation. All models are trained for 10 epochs using the Adam optimizer~\cite{kingma2014adam} with an initial learning rate of $0.0001$. In addition, training of both the unsupervised VAE and the planar NF is performed using the `annealing trick'~\citep{rezende2015variational, van2018sylvester} where the KL divergence term in the loss is scaled by a factor of $\beta$, which increases linearly from 0 to 1 with the number of epochs. We observed that using the `annealing trick' resulted in a significantly lower MSE loss and similar KL loss on the test subset than using the standard ELBO objective \eqref{eq:VAE}.

\paragraph{Dimensionality of the latent space}

\begin{figure}[ht!]
        \includegraphics[width=0.48\linewidth]{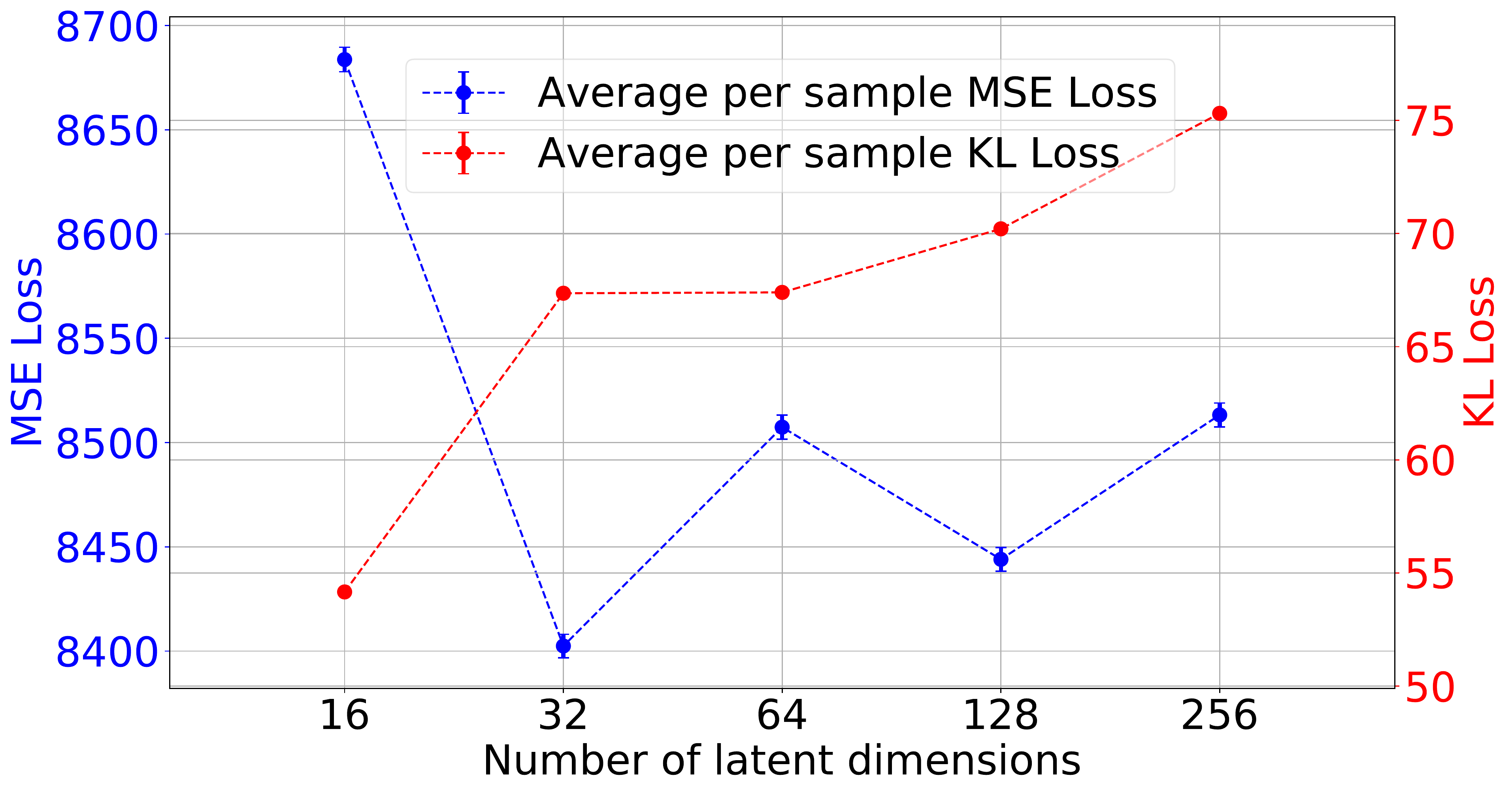}
        \hspace{1mm}
        \includegraphics[width=0.48\linewidth]{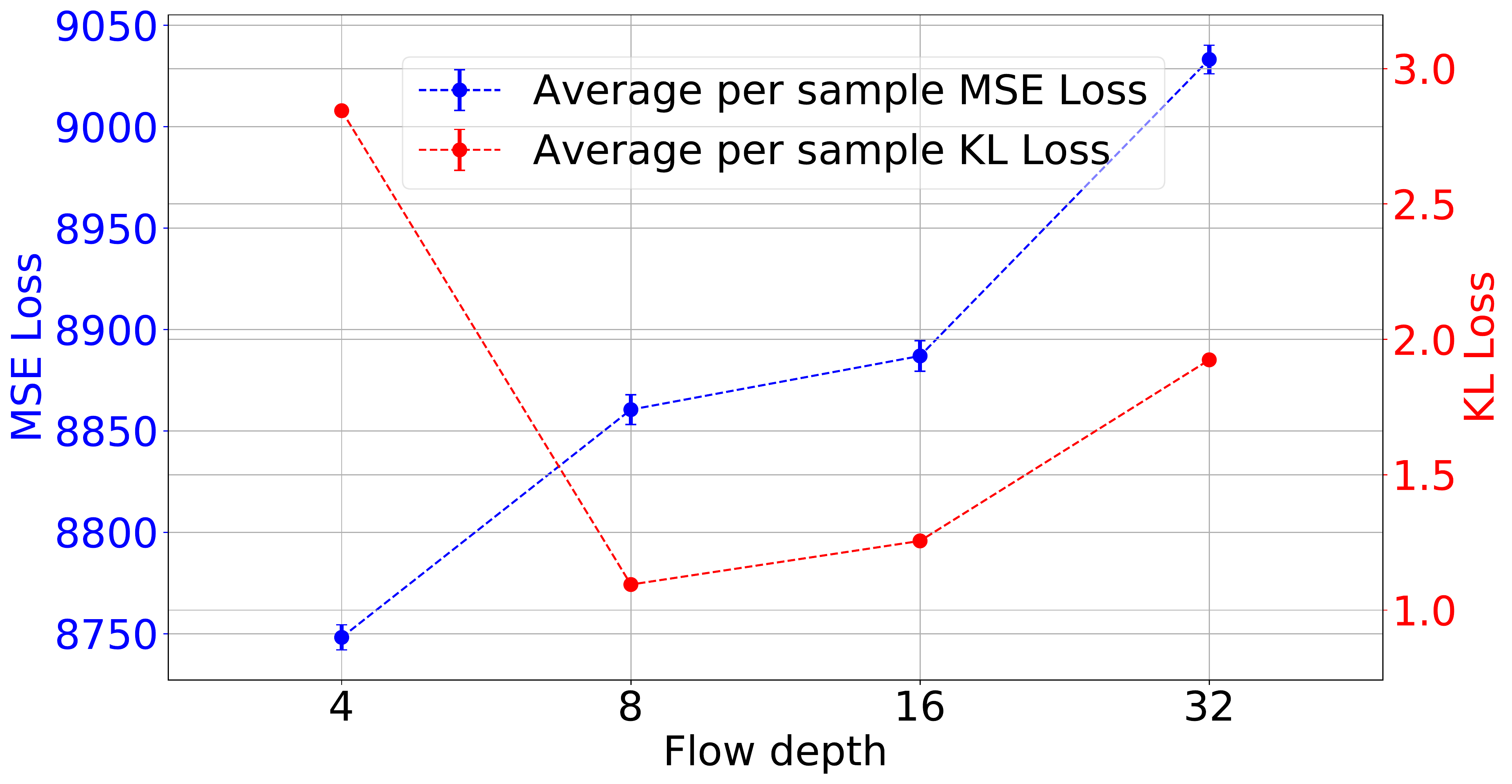}
        \caption{Comparing VAEs with different number of latent dimensions (left) and planar normalizing flows with varying flow depth (right) using MSE and KL loss on the test subset. MSE loss is measured in arbitrary units equivalent to the number of photoelectrons produced at the PMT photocathode and KL loss is measured in nats.\vspace{-0.05in}}
        \label{fig:ld_vs_loss_and_fd_vs_loss}
    \end{figure}

    We tested the VAE with varying number of latent dimensions in order to empirically determine the optimal setting. As shown in Figure \ref{fig:ld_vs_loss_and_fd_vs_loss}, the lowest MSE loss was obtained using 32 latent dimensions. However, using 128 latent dimensions, we achieved better performance in terms of MSE loss and cross entropy loss in the cases of normalizing flows and semi-supervised learning respectively.
	
\paragraph{Planar Normalizing Flows}
	
	We employed planar normalizing flows (Eq.~\ref{eq:PNF}) with a similar amortization strategy to~\citep{rezende2015variational, van2018sylvester} where the flow parameters are considered to be functions of the input rather than part of the global network parameters. As shown in Figure~\ref{fig:ld_vs_loss_and_fd_vs_loss}, we found that using planar normalizing flows, the average KL divergence loss is an order lower than the VAE which can be attributed to having a more flexible posterior distribution. However, the average MSE loss does not improve and in fact is worse than the standard VAE.
    
\paragraph{Event reconstruction and synthetic sample generation}
    
\begin{figure}[ht!]
        \includegraphics[width=\linewidth]{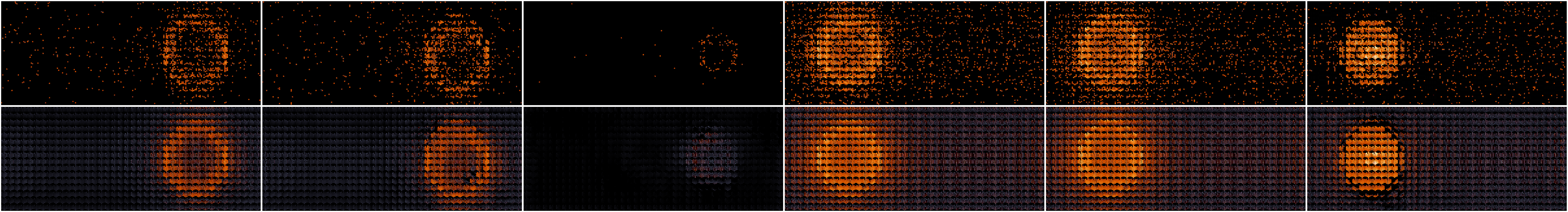}
        \caption{Cherenkov ring images comparing actual simulated events (top) with their corresponding VAE reconstructed events (bottom).\vspace{-0.10in}}
        \label{fig:vae_actual_vs_recon}
    \end{figure}

\begin{figure}[ht!]
        \includegraphics[width=\linewidth]{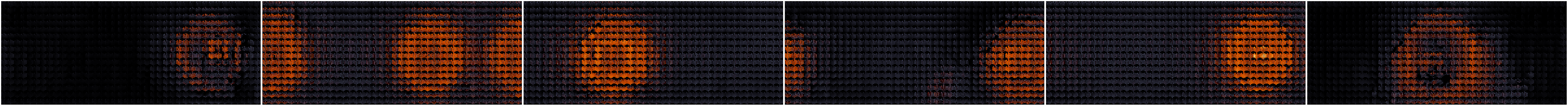}
        \caption{Cherenkov ring images for events randomly sampled from the latent prior p(z) = $\mathcal{N}(0,I)$.\vspace{-0.10in}}
        \label{fig:vae_samples}
    \end{figure}

    The comparison of simulated events and their reconstructions (Figure~\ref{fig:vae_actual_vs_recon}) shows that the VAE is able to capture important features of the Cherenkov ring such as its position, shape, and size. However, the reconstructions show poor ring sharpness and lack of isolated channel charge deposits from PMT dark noise and scattered/reflected light. This is expected due to the limited capacity of the encoder and decoder used in addition to the mean field approximation of the VAE. Images based on sampling from the prior $p(\bz)$ (Figure~\ref{fig:vae_samples}) show similar deficiencies and occasional artifacts, most likely due to prior-posterior divergence.
    
\paragraph{Latent space interpolation}
    In order to perform interpolation in the latent space along the axis of some physically meaningful quantities, we used the k-nearest neighbors algorithm in the feature space of the event azimuthal angle and energy. The latent vectors for the 256 nearest neighbors to some reference point in the feature space were used to find the start and end positions in the latent space. We observed (Figure 4) that linear interpolation along the energy axis yields smooth transitions from one step to the next, however interpolation along the azimuthal angle axis does not. This suggests that not all high level features correspond linearly to directions in the latent space.
    
    \begin{figure}[ht!]
        \includegraphics[width=\linewidth]{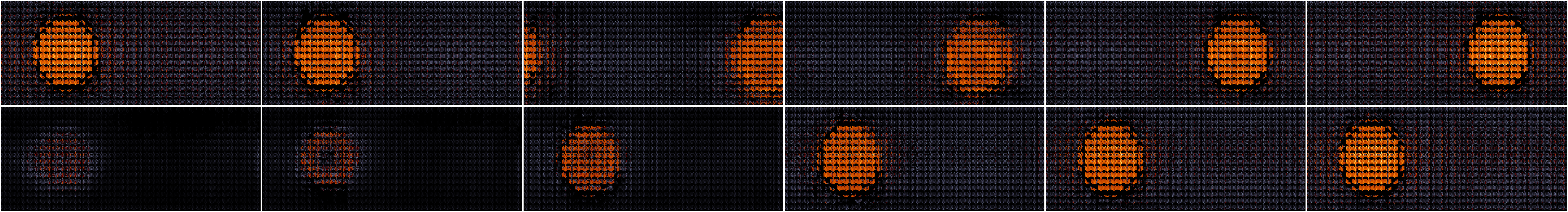}
        \label{fig:linear_interpolation}
        \vspace{-0.15in}
        \caption{Linear interpolation in the latent space for $\mu^-$ events along the angle axis from $\phi = 0$ to $-\pi$ (top) and energy axis from 200 MeV to 800 MeV (bottom).\vspace{-0.25in}}
    \end{figure}

\paragraph{Semi-supervised learning}

	Inspired by the \textbf{M1} strategy of~\citep{kingma2014semi}, we pre-trained the VAE (128 latent dimensions) using approximately 7.2 million training examples and subsequently trained a 4-layer multilayer perceptron (MLP) that takes the deterministic output of the encoder (before the reparameterization) as input. We benchmarked this semi-supervised model against a fully supervised CNN with a similar architecture. The fully supervised CNN and the MLP classifiers were trained for 10 epochs using the Adam optimizer with varying number of labelled examples. We found that in the low labelled data regime, the semi-supervised model consistently outperformed the fully-supervised model and achieved state-of-the-art performance for the task of $\gamma, e^{-}\hspace{1mm}$ versus $\mu^{-}$ event discrimination. As shown in Table~\ref{tbl:dataset_size_vs_classifier_performance}, the \textbf{SS-CNN} model also outperforms the supervised model on $\gamma$ versus $e^-$ classification, considered unfeasible with existing likelihood ratio approaches~\cite{T2K:NCgamma}.
	
	\begin{table}[ht!]
\vspace{-0.1in}
        \centering
		\caption{Comparing the semi-supervised model performance to that of the fully supervised model for various labelled dataset sizes.\newline}
        \label{tbl:dataset_size_vs_classifier_performance}
        \begin{tabular}{lllll}
			\toprule
			
		    \specialcellbold{Number of\\training examples}&
            \multicolumn{2}{c}{\specialcellbold{$\gamma$ background rejection ($\mathbf{\%}$)\\at $\mathbf{50\%}$ $e^-$ signal efficiency}}&
            \multicolumn{2}{c}{\specialcellbold{$\gamma$ background rejection ($\mathbf{\%}$)\\at $\mathbf{80\%}$ $e^-$ signal efficiency}}\\
           
            \cmidrule(lr){2-3}
            \cmidrule(lr){4-5}
           
            &
            
            \specialcellbold{SS-CNN}&
            \specialcellbold{CNN}&
            \specialcellbold{SS-CNN}&
            \specialcellbold{CNN}\\
           
            \midrule
           
           	$11,250$ & $\mathbf{77.6}$ & $76.4$ & $\mathbf{50.7}$ & $46.3$\\
           	$22,500$ & $\mathbf{80.4}$ & $78.1$ & $\mathbf{54.3}$ & $48.5$ \\
           	$45,000$ & $\mathbf{80.7}$ & $79.4$ & $\mathbf{55.9}$ & $49.9$ \\
           	
           \bottomrule
		\end{tabular}
\vspace{-0.15in}
	\end{table}
	
\section{Conclusion}
	We demonstrated the ability of VAEs and NFs to approximate the generative distribution of simulated water Cherenkov detector data, with NFs showing no significant improvement over the standard VAE. When used for the task of classification, the semi-supervised approach demonstrated performance gains over a fully supervised model comparable to those demonstrated in other domains. Reconstruction and synthetic data generation is possible, however the presence of artifacts suggests an improved design of the loss function and/or the generative model is needed. Linear interpolation along the  energy axis displays a smooth behaviour, but along the azimuthal angle axis does not suggesting that more sophisticated interpolation methods may be needed. Through this work, we show the promise of applying generative models to address key challenges of neutrino oscillation experiments such as $\gamma$ vs $e^{-}$ classification and mitigation of experimental uncertainties through future work.

\section*{Acknowledgements}
    The authors would like to thank Akira Konaka, Mark Hartz and Olivia Di Matteo for helpful discussions and feedback on a draft of this paper. This research was enabled in part by support provided by TRIUMF, NSERC and Compute Canada (www.computecanada.ca).
    
\bibliography{paper}

\end{document}